\documentclass[lettersize,journal]{IEEEtran}
\usepackage{amsmath,amsfonts}
\usepackage{algorithmic}
\usepackage{algorithm}
\usepackage{array}
\usepackage[caption=false,font=normalsize,labelfont=sf,textfont=sf]{subfig}
\usepackage{textcomp}
\usepackage{stfloats}
\usepackage{url}
\usepackage{verbatim}
\usepackage{graphicx}
\usepackage{color}
\usepackage{makecell}
\usepackage{multirow}
\usepackage{booktabs}
\usepackage{threeparttable}
\usepackage[colorlinks,
            linkcolor=blue,
            anchorcolor=blue,
            citecolor=blue
            ]{hyperref}
\begin{document}

\title{Calibration of 3D Single-pixel Imaging Systems with a Calibration Field}

\author{Xinyue Ma and Chenxing Wang, \textit{Member, IEEE}
\thanks{This work was supported by the National Natural Science Foundation of China (61828501). (Corresponding author: Chenxing Wang.)}
\thanks{The authors are with the School of Automation, the Key Laboratory of Measurement and Control of Complex Systems of Engineering, Ministry of Education, Southeast University, Nanjing 210096, China
 (e-mail: xymanlfd@seu.edu.cn; cxwang@seu.edu.cn).}}

\markboth{Journal of \LaTeX\ Class Files,~Vol.~x, No.~x, August~2024}%
{Shell \MakeLowercase{\textit{et al.}}: A Sample Article Using IEEEtran.cls for IEEE Journals}


\maketitle

\begin{abstract}
3D single-pixel imaging (SPI) is a promising imaging technique that can be flexibly applied to various wavebands. The main challenge in 3D SPI is that the calibration usually requires a large number of standard points as references, which are tricky to capture using single-pixel detectors. Conventional solutions involve sophisticated device deployment and cumbersome operations, resulting in hundreds of images needed for calibration. In our work, we construct a Calibration Field (CaliF) to efficiently generate the standard points from one single image. A high accuracy of the CaliF is guaranteed by the technique of deep learning and digital twin. We perform experiments with our new method to verify its validity and accuracy. We believe our work holds great potential in 3D SPI systems or even general imaging systems. 
\end{abstract}


\section{Introduction}
\IEEEPARstart{S}{ingle}-pixel imaging (SPI) is a computational imaging technique where sequential patterns are projected to the object and the corresponding transmitted or reflected light intensities are detected by a single-pixel detector (SPD). Then, an object image can be reconstructed from the light intensities by some algorithms. The SPD which is a single photodetector lacks of spatial resolution but has the advantages of wide bandwidth, low dark noise, and high quantum efficiency. Therefore, SPI is widely explored and deeply studied in various fields, such as terahertz imaging \textcolor{blue}{\cite{ref1,ref2}}, infrared imaging \textcolor{blue}{\cite{ref3,ref4}}, X-ray imaging \textcolor{blue}{\cite{ref5}}, imaging in turbid media \textcolor{blue}{\cite{ref6}} and low light imaging \textcolor{blue}{\cite{ref7,ref8}}, etc.

As the demand for 3D imaging is increasing, 3D SPI has shown fast development in recent years. There are mainly three schemes for 3D SPI, i.e., time-of-flight (ToF) \textcolor{blue}{\cite{ref9}}, photometric stereo \textcolor{blue}{\cite{ref10,ref11}}, and stereo vision \textcolor{blue}{\cite{ref4,ref12,ref13,ref14,ref15,ref16,ref17,ref18,ref19,ref20,ref21}}. For ToF, a pulsed laser is emitted to a scene and a high-speed detector receives the reflected photon correspondingly, then the depth can be calculated through the time difference. This scheme has been applied in practice \textcolor{blue}{\cite{ref9, ref22}}, but the precision is limited. The other two schemes are combinations of SPI and computer vision techniques. For SPI with photometric stereo, several SPDs set at different positions detect several light intensity serials separately. These serials are processed to reconstruct the 3D scene based on the technique of shape from shadow. However, depending on the natural reflectivity of objects, the accuracy of this scheme is influenced by the characteristics of objects and the uniformity of illuminations \textcolor{blue}{\cite{ref12}}, and the system deployment is complicated as multiple SPDs are required. For SPI with stereo vision, the number of SPDs is reduced by introducing the structured light technique. It actively creates texture features on the surface of objects, thus achieving higher precision even when the object surface is complex or less textured. Thus, this scheme has been adopted more and more in recent years.

For stereo vision based 3D SPI, calibration is essential if a point cloud is desired. Like conventional 3D imaging techniques \textcolor{blue}{\cite{ref23,ref24,ref25}}, the calibration of 3D SPI mainly includes two strategies. One is to model the relationship of image points to the depth directly. The other is to model the relationship of different sensors based on the triangular stereo method. The former strategy has been the mainstream in 3D SPI \textcolor{blue}{\cite{ref12,ref13,ref14,ref15,ref16}} for a long time since an SPD is difficult to calibrate. However, this strategy requires deploying sophisticated devices and performing tedious operations to precisely control the displacement of the calibration object. Liu et al. \textcolor{blue}{\cite{ref26}} establish a mapping between phase and 3D coordinates by calculating the equations of rays, without using sophisticated devices. However, tedious operations are still required because multiple images need to be captured for calibration.

In contrast, the triangular stereo strategy is more flexible \textcolor{blue}{\cite{ref24,ref25}}. This strategy is to find the matching point pairs between 3D standard points in space and 2D image points to calibrate the intrinsic parameters and extrinsic parameters of imaging devices. To achieve this, there are some self-calibration methods that utilize the geometric consistency constraint relationship \textcolor{blue}{\cite{ref27}} of corresponding image points in multiple frames. They are flexible but have poor robustness and low accuracy because passively extracting the standard points from unknown frames may face many uncertain factors and lead to unstable accuracy. Therefore, calibration based on standard objects is the mainstream since the priori standard information ensures the robustness and accuracy of calibration. With this understanding, the design of standard objects has become a research focus, which has been divided into 1D \textcolor{blue}{\cite{ref28}}, 2D \textcolor{blue}{\cite{ref29}} and 3D \textcolor{blue}{\cite{ref30}} standard objects considering the dimension that the standard points can represent. 2D objects carrying coded patterns are widely used for their high accuracy, better convenience, and low cost. Zhang \textcolor{blue}{\cite{ref29}} proposes the classic camera calibration method, which requires taking multiple images of a 2D planar calibration object. Later, various coded patterns \textcolor{blue}{\cite{ref31,ref32,ref33}} (Chessboard, Circle, ArUcOmni, etc.) and feature detection algorithms \textcolor{blue}{\cite{ref34,ref35}} are proposed to improve calibration accuracy. However, it usually requires imaging a planar calibration object in at least three postures because the planar calibration object cannot provide 3D standard points. To ensure calibration accuracy, the planar calibration object needs to be placed in additional postures to provide a large number of standard points to solve accurate calibration parameters. For calibration combined with Fringe Projection Profilometry (FPP), Zhang et al. \textcolor{blue}{\cite{ref24}} project several fringe patterns onto the calibration object in each posture to calibrate the camera and the projector. Following that, many works \textcolor{blue}{\cite{ref36,ref37,ref38,ref39,ref40}} are proposed to improve the accuracy or speed of calibration. Unfortunately, the limitation of projecting multiple fringe patterns under multiple calibration object postures is not solved, resulting in a huge number of images that need to be captured for calibration. For 3D SPI, Niu et al. \textcolor{blue}{\cite{ref18}} project sinusoidal fringe patterns to a 2D calibration board, and 296 images are needed to calculate the parameters of the 3D SPI system.

To solve the above issues, we propose to create a Calibration Field (CaliF), through which abundant 3D standard points can be obtained directly from only one fringe image of a 3D calibration object. It allows easy establishment of the matching relation between the 3D standard points and 2D image points. Based on the triangular stereo strategy, we utilize the CaliF to calibrate a 3D SPI system with only one image. The contributions of this paper are summarized as follows:

\begin{itemize}
    \item A general and efficient framework for 3D SPI is proposed, which can complete the calibration with simple hardware deployment and easy operations.
    
    \item The concept of Calibration Field (CaliF) is first proposed, which is constructed by using the techniques of deep learning and digital twin, thus the calibration is ensured high precision and convenience.
\end{itemize}

\section{Methodology}
\subsection{Basic Principle and Overall Framework}\label{A}
\begin{figure}[t]
  \centering
   \includegraphics[width=0.8\linewidth]{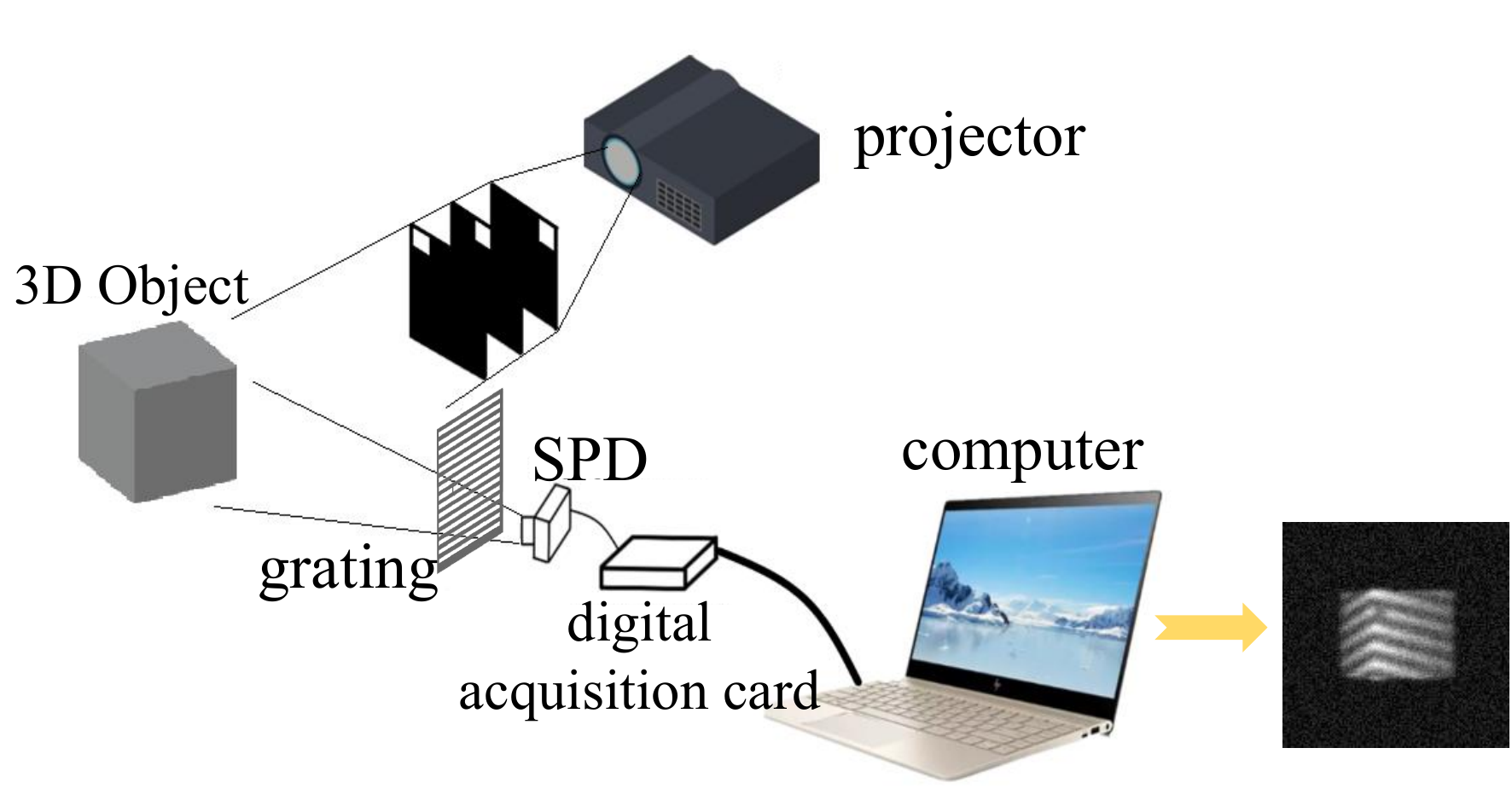}
   \caption{The schematic diagram of a 3D SPI system.}
   \label{Fig.1}
\end{figure}

Figure \textcolor{blue}{\ref{Fig.1}} shows the schematic diagram of a 3D SPI system based on FPP. A projector projects a series of illumination patterns to an object, and the reflected light passes through a grating sheet and is collected by an SPD. Due to the modulation of the grating sheet, a fringe image can be reconstructed from the detected light intensities. Then 3D information can be extracted from the fringe image based on the stereo vision technique.

The light path of the 3D SPI system is a reverse of that of conventional FPP system \textcolor{blue}{\cite{ref13}}. A scene is sampled by the illumination patterns projected by the projector, resembling a “camera” capturing the scene image; the reflected light traverses the grating sheet and then is gathered by the SPD to form fringes, resembling a “projector” projecting a fringe pattern onto the scene. With the pinhole model \textcolor{blue}{\cite{ref18,ref24}}, the imaging of both the “camera” and the “projector” can be described as

\begin{equation}
s\!\!\begin{bmatrix}u \\v \\1\end{bmatrix}\!\!\!=\!\!\!\begin{bmatrix}f^{\mathrm{u}}  &\!\!\!\!\gamma  &\!\!\!\!u_{0}  \\0  &f^{\mathrm{v}}  &\!\!\!\!v_{0}  \\0  &\!\!\!\!0  &\!\!\!\!1 \end{bmatrix}\!\!\!\!\begin{bmatrix}r^{11}  &\!\!\!\!r^{12}  &\!\!\!\!r^{13}  &\!\!\!\!t^{1} \\r^{21}  &\!\!\!\!r^{22}  &\!\!\!\!r^{23}  &\!\!\!\!t^{2} \\r^{31}  &\!\!\!\!r^{32}  &\!\!\!\!r^{33}  &\!\!\!\!t^{3}\end{bmatrix}\!\!\!\!\begin{bmatrix}x_{\mathrm{w}} \\y_{\mathrm{w}} \\z_{\mathrm{w}} \\1\end{bmatrix}\!\!\!=\!\!\!\boldsymbol A[\boldsymbol R\;\boldsymbol t]\!\!\!\begin{bmatrix}x_{\mathrm{w}} \\y_{\mathrm{w}} \\z_{\mathrm{w}} \\1\end{bmatrix},
  \label{eq:1}
\end{equation}
where $s$ is scaling factor, $(u, v)$ and $(x_{\text{w}}, y_{\text{w}}, z_{\text{w}})$ are image coordinate and world coordinate respectively, $f^{\text{u}}$, $f^{\text{v}}$ are focal lengths along the $u$ and $v$ axes of the image plane, $\gamma$ is skew factor of two image axes, $(u_0, v_0)$ is the principal point of the image plane, and these factors constitute the intrinsic parameter matrix denoted as \textbf{\emph{A}}; $r^{ij}$ and $t^i$ represent the rotation and translation parameters, respectively, forming the extrinsic parameter matrix denoted as [\textbf{\emph{R t}}]. Further, \textcolor{blue}{\eqref{eq:1}} can be simplified as the following:

\begin{equation}
s\!\begin{bmatrix}u \\v \\1\end{bmatrix}\!\!=\!\!\boldsymbol M\!\begin{bmatrix}x_{\mathrm{w}} \\y_{\mathrm{w}} \\z_{\mathrm{w}} \\1\end{bmatrix}\!\!=\!\!\begin{bmatrix}m^{11}  &\!\!\!\!m^{12}  &\!\!\!\!m^{13}  &\!\!\!\!m^{14} \\m^{21}  &\!\!\!\!m^{22}  &\!\!\!\!m^{23}  &\!\!\!\!m^{24} \\m^{31}  &\!\!\!\!m^{32}  &\!\!\!\!m^{33}  &\!\!\!\!m^{34}\end{bmatrix}\!\!\!\begin{bmatrix}x_{\mathrm{w}} \\y_{\mathrm{w}} \\z_{\mathrm{w}} \\1\end{bmatrix},
  \label{eq:2}
\end{equation}
where \textbf{\emph{M}}=\textbf{\emph{A}}[\textbf{\emph{R t}}] represents the combined projection matrix. The goal of calibration is to solve the \textbf{\emph{M}} for the “camera” and the “projector”, respectively. Then, spatial points with $(x_{\text{w}}, y_{\text{w}}, z_{\text{w}})$ of any object can be obtained from the image coordinate and the solved \textbf{\emph{M}} of the same devices with the same settings.

In the 3D SPI system in Fig. \textcolor{blue}{\ref{Fig.1}}, the projector plays the role of “camera”, and the SPD combining the grating sheet (SPDG) plays the role of “projector” \textcolor{blue}{\cite{ref13}}. The imaging model of these two are written as

\begin{equation}
s_{\mathrm{p}}\!\begin{bmatrix}u_{\mathrm{p}} \\v_{\mathrm{p}} \\1\end{bmatrix}\!\!=\!\!\boldsymbol M_{\mathrm{p}}\!\begin{bmatrix}x_{\mathrm{w}} \\y_{\mathrm{w}} \\z_{\mathrm{w}} \\1\end{bmatrix}\!\!=\!\!\begin{bmatrix}m_{\mathrm{p}}^{11}  &\!\!\!\!m_{\mathrm{p}}^{12}  &\!\!\!\!m_{\mathrm{p}}^{13}  &\!\!\!\!m_{\mathrm{p}}^{14} \\m_{\mathrm{p}}^{21}  &\!\!\!\!m_{\mathrm{p}}^{22}  &\!\!\!\!m_{\mathrm{p}}^{23}  &\!\!\!\!m_{\mathrm{p}}^{24} \\m_{\mathrm{p}}^{31}  &\!\!\!\!m_{\mathrm{p}}^{32}  &\!\!\!\!m_{\mathrm{p}}^{33}  &\!\!\!\!m_{\mathrm{p}}^{34}\end{bmatrix}\!\!\!\begin{bmatrix}x_{\mathrm{w}} \\y_{\mathrm{w}} \\z_{\mathrm{w}} \\1\end{bmatrix},
  \label{eq:3}
\end{equation}

\begin{equation}
s_{\mathrm{s}}\!\begin{bmatrix}u_{\mathrm{s}} \\v_{\mathrm{s}} \\1\end{bmatrix}\!\!=\!\!\boldsymbol M_{\mathrm{s}}\!\begin{bmatrix}x_{\mathrm{w}} \\y_{\mathrm{w}} \\z_{\mathrm{w}} \\1\end{bmatrix}\!\!=\!\!\begin{bmatrix}m_{\mathrm{s}}^{11}  &\!\!\!\!m_{\mathrm{s}}^{12}  &\!\!\!\!m_{\mathrm{s}}^{13}  &\!\!\!\!m_{\mathrm{s}}^{14} \\m_{\mathrm{s}}^{21}  &\!\!\!\!m_{\mathrm{s}}^{22}  &\!\!\!\!m_{\mathrm{s}}^{23}  &\!\!\!\!m_{\mathrm{s}}^{24} \\m_{\mathrm{s}}^{31}  &\!\!\!\!m_{\mathrm{s}}^{32}  &\!\!\!\!m_{\mathrm{s}}^{33}  &\!\!\!\!m_{\mathrm{s}}^{34}\end{bmatrix}\!\!\!\begin{bmatrix}x_{\mathrm{w}} \\y_{\mathrm{w}} \\z_{\mathrm{w}} \\1\end{bmatrix},
  \label{eq:4}
\end{equation}
where the subscript p and s indicate the projector and the SPDG, respectively. Since the projector works like a “camera”, the solution of \textbf{\emph{M}}$_{\text{p}}$ is simple and needs only common operations, where the $(u_{\text{p}}, v_{\text{p}})$ can be extracted directly from the captured images. There are 12 variables in the \textbf{\emph{M}}$_{\text{p}}$, thus at least 6 pairs of matching points, $(x_{\text{w}}, y_{\text{w}}, z_{\text{w}})$-$(u_{\text{p}}, v_{\text{p}})$, are required to solve the \textbf{\emph{M}}$_{\text{p}}$. If a planar calibration object is used, the \textbf{\emph{M}}$_{\text{p}}$ cannot be solved directly. The reason is that there is $z\equiv0$ for the points on the plane, so the parameters of the third column of \textbf{\emph{M}}$_{\text{p}}$ cannot be solved. Hence, the intrinsic parameter and the extrinsic parameter must be solved separately through homography matrices \textcolor{blue}{\cite{ref29}}, which requires that the planar calibration object should be placed in at least 3 postures.

Solving \textbf{\emph{M}}$_{\text{s}}$ is difficult because the SPDG works like a “projector”, so $(u_{\text{s}}, v_{\text{s}})$ cannot be obtained directly like capturing images using a camera. In FPP \textcolor{blue}{\cite{ref24}}, to cope with this issue, a point $(u_{\text{s}}, v_{\text{s}})$ is matched to a corresponding pixel point $(u_{\text{p}}, v_{\text{p}})$ by the following model:

\begin{equation}
\left\{\begin{aligned}u_\text{s} =\boldsymbol \phi _\text{v}(u_\text{p},v_\text{p})\times T/2\pi \\v_\text{s} =\boldsymbol \phi _\text{h}(u_\text{p},v_\text{p})\times T/2\pi\end{aligned}\right.,
  \label{eq:5}
\end{equation}
where $\boldsymbol \phi(u_{\text{p}}, v_{\text{p}})$ denotes the absolute phase value of a fringe image captured after projecting a fringe pattern onto a calibration object, the subscript v and h denote the fringe directions, vertical and horizontal, respectively, and $T$ is the period of fringes. Similar to solving the \textbf{\emph{M}}$_{\text{p}}$, the solution of \textbf{\emph{M}}$_{\text{s}}$ should be done by finding several pairs of matching points ($(x_{\text{w}}, y_{\text{w}}, z_{\text{w}})$-$(u_{\text{s}}, v_{\text{s}})$) with the help of \textcolor{blue}{\eqref{eq:5}}. However, in the 3D SPI system in Fig. \textcolor{blue}{\ref{Fig.1}}, fringes are generated by a grating sheet that has only one direction, thus only one of $u_{\text{s}}$ and $v_{\text{s}}$ can be obtained once the device is deployed. To complete calibration using fringe patterns with a single direction, we refer to \textcolor{blue}{\cite{ref41}} stating that only the corresponding two rows of \textbf{\emph{M}}$_{\text{s}}$ are needed according to the direction of fringes, in the application of calculating $(x_{\text{w}}, y_{\text{w}}, z_{\text{w}})$. Our grating sheet is customized with horizontal grid lines, so horizontal fringes are provided, then only $v_{\text{s}}$ can be obtained, and the first row of \textbf{\emph{M}}$_{\text{s}}$ corresponding to $u_{\text{s}}$ can be omitted, leading to a simplification of \textcolor{blue}{\eqref{eq:4}} as

\begin{equation}
\begin{aligned}
    v_{\mathrm{s}}x_{\mathrm{w}}\frac{m_{\mathrm{s}}^{31}}{m_{\mathrm{s}}^{24}}+v_{\mathrm{s}}y_{\mathrm{w}}\frac{m_{\mathrm{s}}^{32}}{m_{\mathrm{s}}^{24}}+v_{\mathrm{s}}z_{\mathrm{w}}\frac{m_{\mathrm{s}}^{33}}{m_{\mathrm{s}}^{24}}+v_{\mathrm{s}}\frac{m_{\mathrm{s}}^{34}}{m_{\mathrm{s}}^{24}}\\-x_{\mathrm{w}}\frac{m_{\mathrm{s}}^{21}}{m_{\mathrm{s}}^{24}}-y_{\mathrm{w}}\frac{m_{\mathrm{s}}^{22}}{m_{\mathrm{s}}^{24}}-z_{\mathrm{w}}\frac{m_{\mathrm{s}}^{23}}{m_{\mathrm{s}}^{24}}=1.
\end{aligned}
  \label{eq:6}
\end{equation}
\textcolor{blue}{\eqref{eq:6}} is further rewritten as

\begin{equation}
\boldsymbol b \boldsymbol m_{\mathrm{s}}=1,
  \label{eq:7}
\end{equation}
where

\begin{equation}
\boldsymbol b\!=\!\begin{bmatrix}v_{\mathrm{s}}x_{\mathrm{w}}&v_{\mathrm{s}}y_{\mathrm{w}}&v_{\mathrm{s}}z_{\mathrm{w}}&v_{\mathrm{s}}&-x_{\mathrm{w}}&-y_{\mathrm{w}}&-z_{\mathrm{w}}\end{bmatrix},
    \label{eq:8}
\end{equation}

\begin{equation}
\boldsymbol m_{\mathrm{s}}\!=\!\begin{bmatrix}\frac{m_{\mathrm{s}}^{31}}{m_{\mathrm{s}}^{24}}\!   \!\!&\frac{m_{\mathrm{s}}^{32}}{m_{\mathrm{s}}^{24}}\!   \!\!  &\frac{m_{\mathrm{s}}^{33}}{m_{\mathrm{s}}^{24}}\!   \!\!  &\frac{m_{\mathrm{s}}^{34}}{m_{\mathrm{s}}^{24}}\!   \!\!  &\frac{m_{\mathrm{s}}^{21}}{m_{\mathrm{s}}^{24}}\!   \!\!  &\frac{m_{\mathrm{s}}^{22}}{m_{\mathrm{s}}^{24}}\!   \!\!  &\frac{m_{\mathrm{s}}^{23}}{m_{\mathrm{s}}^{24}}\end{bmatrix}^{\mathrm{T} }.
    \label{eq:9}
\end{equation}
The vector \textbf{\emph{m}}$_{\text{s}}$ is a simplification of the \textbf{\emph{M}}$_{\text{s}}$, which shows that the number of unknown parameters is reduced to 7, thus more than 7 pairs of matching points ($(x_{\text{w}}, y_{\text{w}}, z_{\text{w}})$-$v_{\text{s}}$) are required to solve the \textbf{\emph{m}}$_{\text{s}}$.

With \textbf{\emph{M}}$_{\text{p}}$ and \textbf{\emph{m}}$_{\text{s}}$ being solved, 3D measurement can be applied by combining \textcolor{blue}{\eqref{eq:3}} and \textcolor{blue}{\eqref{eq:6}} \textcolor{blue}{\cite{ref41}}:

\begin{equation}
\begin{bmatrix}x_{\mathrm{w}}   &y_{\mathrm{w}}  &z_{\mathrm{w}}\end{bmatrix}^{\mathrm{T}}=\boldsymbol C^{-1} \boldsymbol d,
    \label{eq:10}
\end{equation}

\begin{equation}
\boldsymbol C\!=\!\begin{bmatrix}m_{\mathrm{p}}^{11}\!-\!u_{\mathrm{p}}m_{\mathrm{p}}^{31}  &m_{\mathrm{p}}^{12}\!-\!u_{\mathrm{p}}m_{\mathrm{p}}^{32}  &m_{\mathrm{p}}^{13}\!-\!u_{\mathrm{p}}m_{\mathrm{p}}^{33}\\m_{\mathrm{p}}^{21}\!-\!v_{\mathrm{p}}m_{\mathrm{p}}^{31}  &m_{\mathrm{p}}^{22}\!-\!v_{\mathrm{p}}m_{\mathrm{p}}^{32}  &m_{\mathrm{p}}^{23}\!-\!v_{\mathrm{p}}m_{\mathrm{p}}^{33} \\\frac{m_{\mathrm{s}}^{21}}{m_{\mathrm{s}}^{24}}\!- \!v_{\mathrm{s}}\frac{m_{\mathrm{s}}^{31}}{m_{\mathrm{s}}^{24}}  &\frac{m_{\mathrm{s}}^{22}}{m_{\mathrm{s}}^{24}}\!-\! v_{\mathrm{s}}\frac{m_{\mathrm{s}}^{32}}{m_{\mathrm{s}}^{24}}  &\frac{m_{\mathrm{s}}^{23}}{m_{\mathrm{s}}^{24}}\!-\! v_{\mathrm{s}}\frac{m_{\mathrm{s}}^{33}}{m_{\mathrm{s}}^{24}}\end{bmatrix},
    \label{eq:11}
\end{equation}

\begin{equation}
\boldsymbol d=\begin{bmatrix}u_{\mathrm{p}}m_{\mathrm{p}}^{34}-m_{\mathrm{p}}^{14}  &v_{\mathrm{p}}m_{\mathrm{p}}^{34}-m_{\mathrm{p}}^{24}  &v_{\mathrm{s}}\frac{m_{\mathrm{s}}^{34}}{m_{\mathrm{s}}^{24}}-1 \end{bmatrix}^{\mathrm{T}}.
    \label{eq:12}
\end{equation}

According to the analysis above, the calibration of the 3D SPI system is to solve the \textbf{\emph{M}}$_{\text{p}}$ and \textbf{\emph{m}}$_{\text{s}}$ by finding sufficient matching point pairs. However, this faces challenges below.

\begin{itemize}
    \item To reconstruct an image to obtain the $(u_{\text{p}}, v_{\text{p}})$ needs huge sampling and detecting operations for the SPI system, due to the lack of spatial resolution for SPD.
    
    \item To calibrate an imaging device using a planar calibration object requires that the plane must be placed in multiple postures according to Zhang’s method \textcolor{blue}{\cite{ref29}}, resulting in more images being needed.

    \item To obtain $v_{\text{s}}$, the absolute phase needs to be solved from the fringe image, and the phase-shifting method is usually used for higher accuracy and robustness, which needs many fringe images with phase shifts.
\end{itemize}

Considering above requirements as a whole, the number of required images should increase many times, while the operations will increase explosively due to using SPD. For SPI systems, imaging with multiple sampling and detecting operations is unavoidable, but reducing the number of images is possible. Therefore, in this paper, we explore these ways:

\begin{itemize}
    \item A standard white cube with a simple structure and low cost is used as the calibration object, which can provide a number of 3D standard points by imaging it once.
    
    \item A Calibration Field (CaliF) is proposed and constructed, which has the ability to generate a pointmap containing abundant 3D standard points of the cube to solve the \textbf{\emph{M}}$_{\text{p}}$ and \textbf{\emph{m}}$_{\text{s}}$ from only one cube fringe image.

    \item A Fringe Analysis Module (FriAM) is constructed to obtain the absolute phase for obtaining $v_{\text{s}}$, also needing only one fringe image.
\end{itemize}

\begin{figure}[t]
  \centering
   \includegraphics[width=1.0\linewidth]{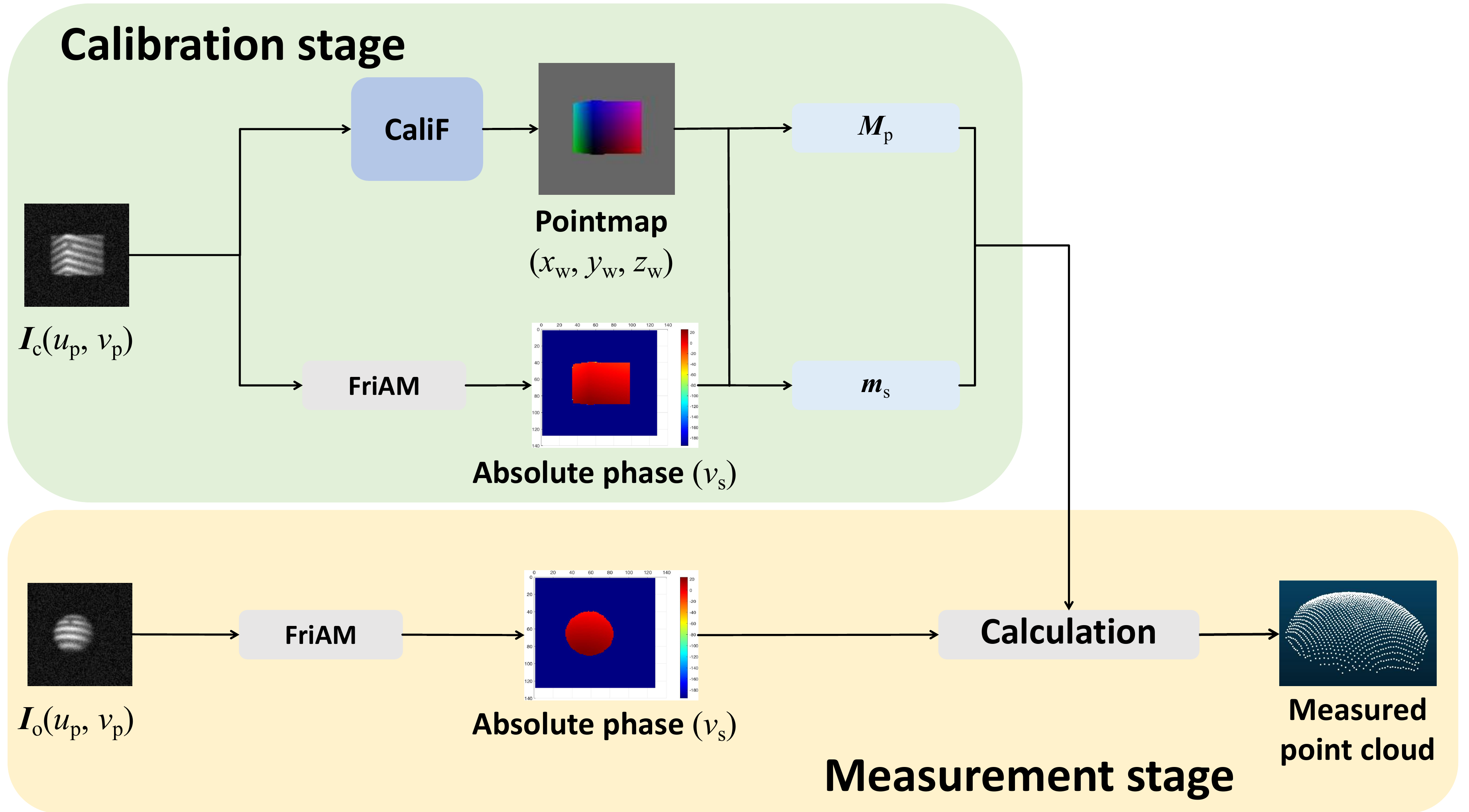}
   \caption{Framework of the proposed method.}
   \label{Fig.2}
\end{figure}

Figure \textcolor{blue}{\ref{Fig.2}} shows the framework of our method, including a calibration stage and a measurement stage. A fringe image of a white cube or a measured object, denoted as \textbf{\emph{I}}$_{\text{c}}\in\mathbb{R}^{W\times H\times1}$ or \textbf{\emph{I}}$_{\text{o}}\in\mathbb{R}^{W\times H\times1}$, can be obtained using the system in Fig. \textcolor{blue}{\ref{Fig.1}}, using the method in \textcolor{blue}{\cite{ref21}}. Then, \textbf{\emph{I}}$_{\text{c}}$ is sent to the CaliF and the FriAM meantime to obtain \textbf{\emph{M}}$_{\text{p}}$ and \textbf{\emph{m}}$_{\text{s}}$, and \textbf{\emph{I}}$_{\text{o}}$ is input into the measurement stage to calculate absolute phase and finally to obtain the point cloud.

\subsection{The Calibration Field}
To solve the \textbf{\emph{M}}$_{\text{p}}$ and \textbf{\emph{m}}$_{\text{s}}$ in \textcolor{blue}{\eqref{eq:3}} and \textcolor{blue}{\eqref{eq:9}}, abundant accurate matching point pairs, $(x_{\text{w}}, y_{\text{w}}, z_{\text{w}})$-$(u_{\text{p}}, v_{\text{p}})$ and $(x_{\text{w}}, y_{\text{w}}, z_{\text{w}})$-$v_{\text{s}}$, are required. If a planar calibration object is used, it must be moved many times to solve the calibration parameters, because the plane does not have 3D points \textcolor{blue}{\cite{ref29}}, as analyzed above. To simplify the operation, we apply a standard cube to provide the required 3D standard points by imaging it once, and then the solutions of \textbf{\emph{M}}$_{\text{p}}$ and \textbf{\emph{m}}$_{\text{s}}$ can be conducted from the matching points directly. Hence, the key for calibration is to precisely extract the matching points.

In traditional methods \textcolor{blue}{\cite{ref42,ref43}}, some coded points are marked on the surface of the calibration object and then are extracted to obtain matching point pairs for solving calibration parameters. These methods face the challenge of accurately extracting the coded points, which is even more difficult if fringes are projected to the calibration object. With a white cube as the calibration object, we propose to construct a CaliF that can output a precise pointmap \textcolor{blue}{\cite{ref44}} from a single-frame fringe image of the cube. The pointmap is defined as a representation similar to an RGB image that is written as \textbf{\emph{I}}$\in\mathbb{R}^{W\times H\times3}$. With R, G, and B values replaced, the three channels of a pointmap \textbf{\emph{P}}$\in\mathbb{R}^{W\times H\times3}$ record the world coordinates of 3D standard points as:

\begin{equation}
\left\{\begin{aligned}\boldsymbol P(u_\text{p},v_\text{p},0)=x_\text{w} \\\boldsymbol P(u_\text{p},v_\text{p},1)=y_\text{w} \\\boldsymbol P(u_\text{p},v_\text{p},2)=z_\text{w}\end{aligned}\right.,u_\text{p}\in \left \{ 1\cdots W \right \},v_\text{p}\in \left \{ 1\cdots H \right \}
  \label{eq:13}
\end{equation}
Therefore, the pointmap can naturally match all image pixels $(u_{\text{p}}, v_{\text{p}})$ to 3D standard points $(x_{\text{w}}, y_{\text{w}}, z_{\text{w}})$ one by one, thus sufficient matching pairs can be obtained for calculating the corresponding projection matrix \textbf{\emph{M}}$_{\text{p}}$.

As is well-known, obtaining 3D information from a single image faces the problem of scale ambiguity. To solve this, our CaliF only targets white cubes with known scales because the size of the calibration object should be known in practice. A large number of pairs of pointmaps and fringe images of the cube constitute the dataset. A deep learning network is trained using this dataset to construct our CaliF. During training, the $x_\text{w}$, $y_\text{w}$, and $z_\text{w}$ coordinates in the pointmap are normalized by dividing them by the side length of the cube. In practice, the CaliF outputs a precise pointmap recording the normalized coordinates, and the absolute 3D coordinates can be obtained by multiplying the normalized coordinates by the side length of the cube.

\begin{figure}[t]
  \centering
   \includegraphics[width=0.8\linewidth]{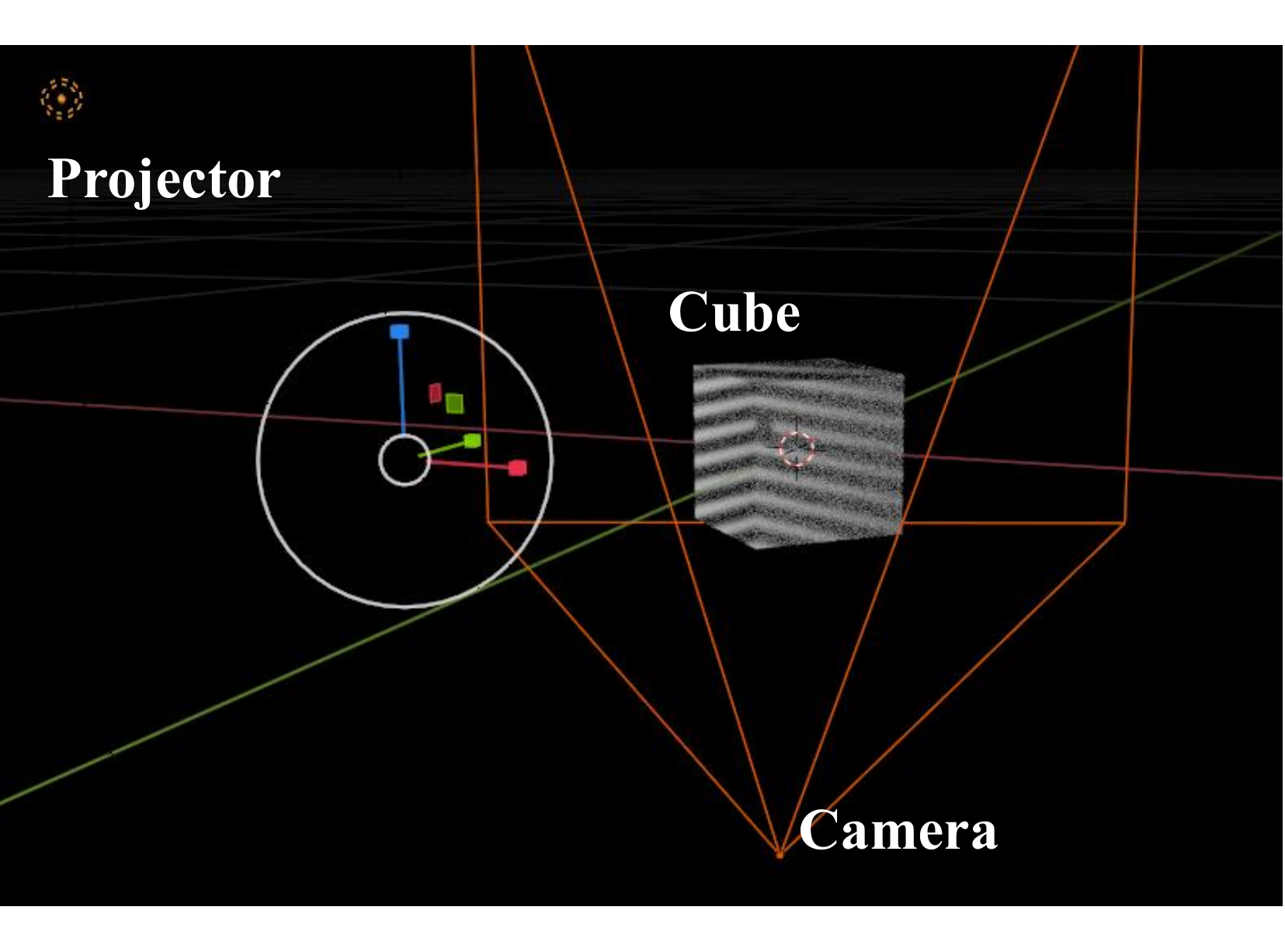}

   \caption{The virtual FPP system.}
   \label{Fig.3}
\end{figure}

In this paper, we adopt the network used in \textcolor{blue}{\cite{ref45}} to serve as the CaliF. The details of the network are available at \textcolor{blue}{\cite{ref46}}. We adopt the digital twin technique \textcolor{blue}{\cite{ref47,ref48}} to generate the training data. A virtual FPP system and a digital twin of a real white cube are set in a software Blender. The virtual FPP system is shown in Fig. \textcolor{blue}{\ref{Fig.3}}, which consists of two devices, a projector (equivalent to the SPDG in Fig. \textcolor{blue}{\ref{Fig.1}}) and a camera (equivalent to the projector in Fig. \textcolor{blue}{\ref{Fig.1}}). With the cube set in the virtual FPP system, a large number of depth maps and fringe images of the cube can be rendered conveniently. Some details of constructing the virtual FPP system can be found in \textcolor{blue}{\cite{ref49}}.

Then, true pointmaps are required as labels to train the network. The generation of a pointmap is to calculate the $(x_{\text{w}}, y_{\text{w}}, z_{\text{w}})$ exactly. Given an image pixel with $(u_{\text{p}}, v_{\text{p}})$, the pixel can be back-projected to the world coordinate system based on \textcolor{blue}{\eqref{eq:1}}:

\begin{equation}
\begin{bmatrix}x_{\mathrm{w}}\\y_{\mathrm{w}}\\z_{\mathrm{w}}\end{bmatrix}=\boldsymbol R ^{-1}(s\boldsymbol A ^{-1}\begin{bmatrix}u_{\mathrm{p}}\\v_{\mathrm{p}}\\1\end{bmatrix}-\boldsymbol t),
    \label{eq:14}
\end{equation}
where \textbf{\emph{R}} and \textbf{\emph{t}} are the rotation and translation matrices separately, \textbf{\emph{A}} is the matrix of intrinsic parameters of the camera, and $s$ is the scaling factor equaling to the depth value of the cube model. These parameters can be set in Blender \textcolor{blue}{\cite{ref49}}. With \textbf{\emph{A}}, \textbf{\emph{R}}, and \textbf{\emph{t}} being diverse, we build a training dataset with fringe images and pointmaps as inputs and labels, respectively. Then a CaliF about the cube is constructed by training the deep learning network using this dataset. In practical applications, if a cube fringe image is captured, which is similar to the one captured in the virtual system, the CaliF can output an accurate pointmap.

\begin{figure}[t]
  \centering
   \includegraphics[width=1.0\linewidth]{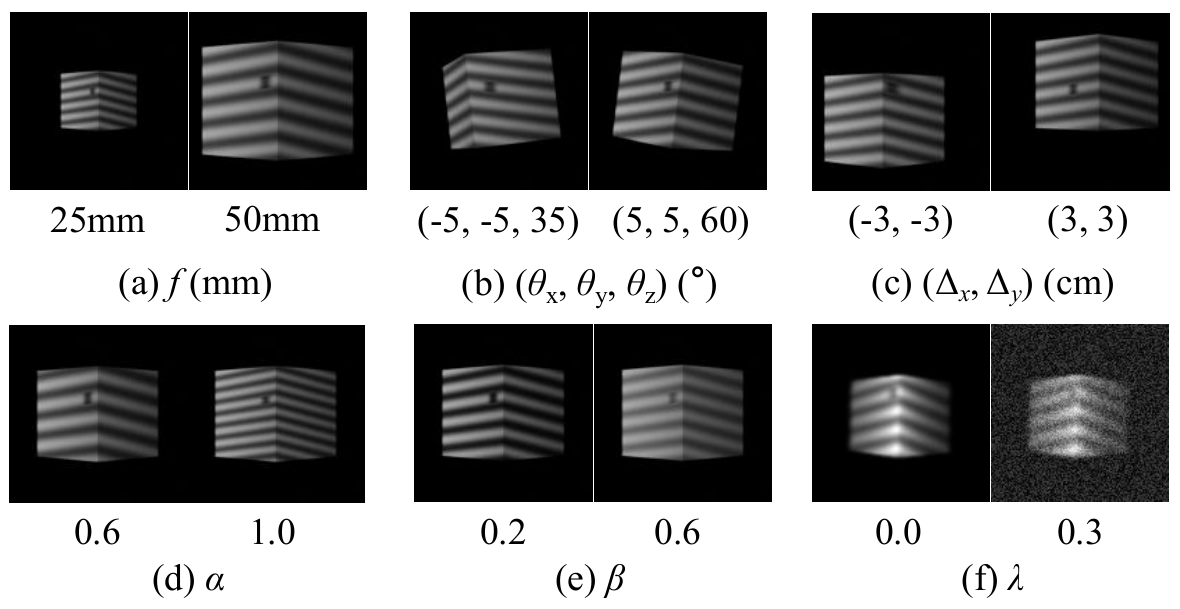}
   \caption{Display of images with different parameters.}
   \label{Fig.4}
\end{figure}

In order to adapt to actual usage situations, the dataset should contain diverse training data by setting the parameters within a range in Blender. The larger the range, the wider the application in practice, while a larger dataset and more advanced network structure are required to train a more powerful CaliF. In this paper, the intrinsic matrix \textbf{\emph{A}} is changed mainly by varying the camera focal length, \textbf{\emph{R}} is changed by rotating the cube within a range, and \textbf{\emph{t}} is changed by shifting the cube model randomly. It is worth noting that at least two surfaces of the cube model should be captured at the same time in an image, to ensure that sufficient matching points embedded with 3D information can be provided. Except for the parameters above, some factors also need to be set in variety for the generalization of the CaliF, such as the period and intensity of fringes, etc. Detailed settings are listed in Table. \textcolor{blue}{\ref{tab:1}} and some examples are shown in Fig. \textcolor{blue}{\ref{Fig.4}}, where Fig. \textcolor{blue}{\ref{Fig.4}}(f) is worth noting, as the image intensity is adjusted with additional consideration, i.e., the brightness of a point on the object is higher if it is relatively closer to the camera, which is more evident for SPI system.

\begin{table*}[!t]
  \caption{Parameter settings for the virtual system.}
  \centering
  \tabcolsep=0.3cm                
  \renewcommand\arraystretch{1.5}   
  \begin{tabular}{cccccccccc}
    \toprule  
    factor & $f$ (mm) & $\theta_\text{x}$ (°) & $\theta_\text{y}$ (°) & $\theta_\text{z}$ (°) & $\Delta_\text{x}$ (cm) & $\Delta_\text{y}$ (cm) & $\alpha$ & $\beta$ & $\lambda$\\
    \midrule
    range & [25, 50] & [-5, 5] & [-5, 5] & [35, 60] & [-3, 3] & [-3, 3] & [0.6, 1.0] & [0.2, 0.6] & [0.0, 0.3]\\
    \bottomrule
  \end{tabular}
  \begin{tablenotes}    
        \footnotesize               
        \item $f$ is the focal length of the camera; $(\theta_\text{x}, \theta_\text{y}, \theta_\text{z})$ are the rotation angles of the cube; $\Delta_\text{x}$ and $\Delta_\text{y}$ are the translations of the cube in the horizontal and vertical direction, respectively; $\alpha$ controls fringe period \textcolor{blue}{\cite{ref49}}; $\beta$ is the valley intensity of fringes with peak intensity as 1; $\lambda$ is the maximum intensity of random noise, relative to the pixel value distributed between 0 and 1.          
  \end{tablenotes}            
  \label{tab:1}
\end{table*}

\subsection{Implementation of Calibration Using the CaliF}
In the practical application, a real white cube is used to conduct the calibration. The 3D printing technique is recommended to produce the cube, which has high precision, high convenience, and low cost. According to the framework in Fig. \textcolor{blue}{\ref{Fig.2}}, the main task in the calibration stage is to solve \textbf{\emph{M}}$_{\text{p}}$ and \textbf{\emph{m}}$_{\text{s}}$. The solution of \textbf{\emph{M}}$_{\text{p}}$ requires at least 6 pairs of 3D-2D standard points ($(x_{\text{w}}, y_{\text{w}}, z_{\text{w}})$-$(u_{\text{p}}, v_{\text{p}})$) as analyzed in Subsection \textcolor{blue}{\ref{A}}, which can be obtained easily in the pointmap output by our CaliF. To ensure precision, we judge the quality of the 3D points from the pointmap by evaluating whether their perpendicular distances to the ideal surface of the cube model are less than 0.01mm. So usually more than 2000 pairs of matching points can be selected. Then, the \textbf{\emph{M}}$_{\text{p}}$ is solved by the means of Singular Value Decomposition (SVD) \textcolor{blue}{\cite{ref50}}.

The solution of \textbf{\emph{m}}$_{\text{s}}$ requires more than 7 pairs of matching points ($(x_{\text{w}}, y_{\text{w}}, z_{\text{w}})$-$v_{\text{s}}$). The matching pairs, $(x_{\text{w}}, y_{\text{w}}, z_{\text{w}})$-$(u_{\text{p}}, v_{\text{p}})$, are obtained using the CaliF, and then we calculate the absolute phase value $\boldsymbol \phi_\text{h} (u_{\text{p}}, v_{\text{p}})$ of the fringe image $\boldsymbol I(u_{\text{p}}, v_{\text{p}})$ to obtain the $v_\text{s}$ by \textcolor{blue}{\eqref{eq:5}}, thus, using $\boldsymbol \phi_\text{h} (u_{\text{p}}, v_{\text{p}})$ as a bridge, $(x_{\text{w}}, y_{\text{w}}, z_{\text{w}})$ and $v_\text{s}$ are matched. Then, the \textbf{\emph{m}}$_{\text{s}}$ is solved by the Least Squares Method \textcolor{blue}{\cite{ref41}}.

\begin{figure*}[t]
  \centering
   \includegraphics[width=1.0\linewidth]{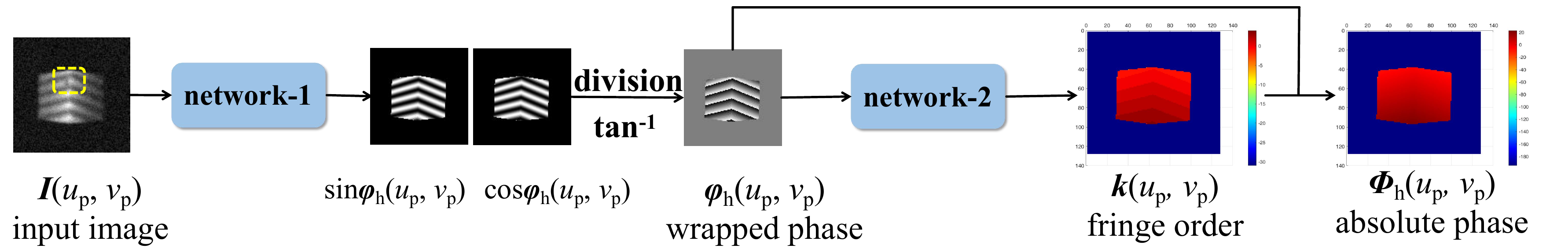}
   \caption{Schematic of the FriAM.}
   \label{Fig.5}
\end{figure*}

To obtain the absolute phase map $\boldsymbol \phi_\text{h}$ from a single fringe image, we construct a FriAM based on deep learning referring to \textcolor{blue}{\cite{ref51}} and \textcolor{blue}{\cite{ref52}}. A network-1 is constructed to simultaneously calculate $\sin\boldsymbol \varphi_\text{h}$ and $\cos\boldsymbol \varphi_\text{h}$ from a fringe image, where $\boldsymbol \varphi_\text{h}$ is a map of wrapped phase. By dividing $\sin\boldsymbol \varphi_\text{h}$ by $\cos\boldsymbol \varphi_\text{h}$ and then calculating the arctangent function, we can obtain the $\boldsymbol \varphi_\text{h}$ that wraps within $[-\pi, \pi]$. The absolute phase map $\boldsymbol \phi_\text{h}$ can be calculated with $\boldsymbol \phi_\text{h}=2\boldsymbol k \pi+\boldsymbol \varphi_\text{h}$ where the fringe order map $\boldsymbol k$ is output by a trained network-2. To avoid ambiguity, we design a marker on the fringe pattern to mark the position of the $0^\text{th}$ order. Figure \textcolor{blue}{\ref{Fig.5}} shows the framework of the FriAM. The marker connects the two black fringes, as indicated in the dashed yellow box on the input image. In the real 3D SPI system, we also make such a marker on the grating sheet to mark the $0^\text{th}$ order. In this paper, the network-1 also adopts the same one used for constructing the CaliF. The network-2 is trained like a segmentation task \textcolor{blue}{\cite{ref52}} using the popular powerful Unet \textcolor{blue}{\cite{ref53}}. The training datasets are also calculated from the fringe images rendered by Blender, combined with the trigonometric function as well as the inverse process of \textcolor{blue}{\eqref{eq:5}}.

As mentioned above, the calibration is completed, and then 3D measurement can be performed by \textcolor{blue}{\eqref{eq:10}}.

\section{Experiments}
\subsection{Simulation}
In this paper, the Adam optimizer and the mean-squared-error (MSE) loss function are adopted to train all networks. The CaliF is trained for 200 epochs, with a learning rate of 0.0005 for the first 100 epochs and 0.0001 for the last 100 epochs. The batch size is 4. In the FriAM, the training parameters of the network-1 are the same as the ones of the CaliF, and the network-2 is trained for 100 epochs with a learning rate of 0.0001 and a batch size of 4.

In this research, two types of datasets are obtained from the virtual system in Blender. One used for training the CaliF consists of 10,000 sets of data of cubes. The other one used for measurement consists of 30,800 sets of data consisting of 154 objects from the Thingi10k \textcolor{blue}{\cite{ref54}}.

\begin{figure}[t]
  \centering
   \includegraphics[width=0.7\linewidth]{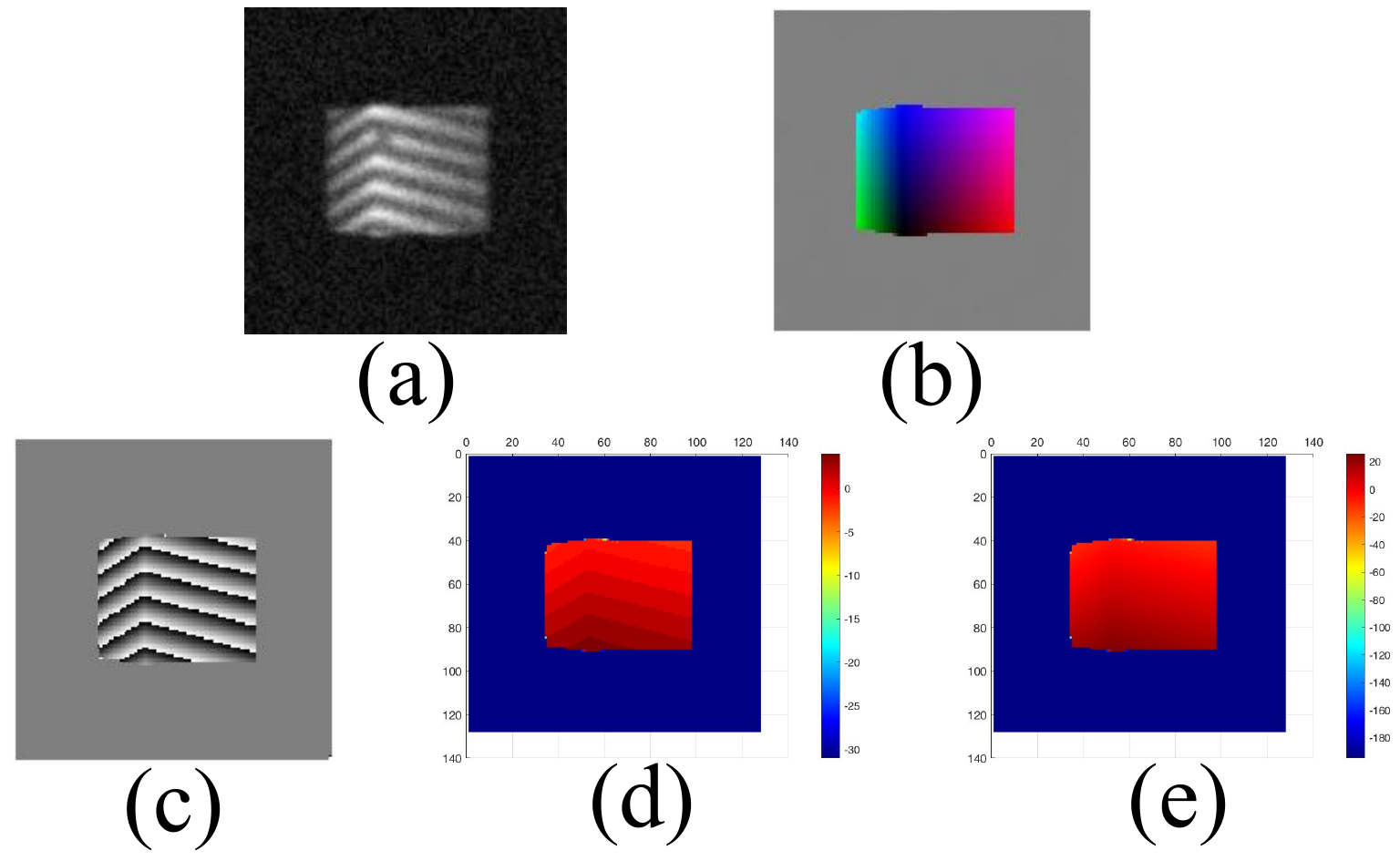}
   \caption{The intermediate results. (a) is the cube fringe image, (b) is the pointmap output by CaliF, (c) is the wrapped phase map obtained from the network-1, (d) is the fringe order map obtained from the network-2, and (e) is the absolute phase map.}
   \label{Fig.6}
\end{figure}

Simulation experiments are performed to verify the accuracy of our method. A fringe image of a cube model is rendered in the virtual system. The side length of the cube is 50mm and the resolution of the rendered image is 128×128. Then, the cube fringe image is sent to the CaliF and the FriAM to calibrate the system. The intermediate results generated during the calibration process are shown in Fig. \textcolor{blue}{\ref{Fig.6}}, where the pointmap and the absolute phase map are used to solve the \textbf{\emph{M}}$_{\text{p}}$ and the \textbf{\emph{m}}$_{\text{s}}$.

\begin{equation}
\boldsymbol M_{\mathrm{p}}=\begin{bmatrix}0.673  &-0.192  &0.003  &0.044 \\0.071  &0.120  &-0.643  &0.045 \\0.129  &0.220  &0.029  &0.089\end{bmatrix},
    \label{eq:15}
\end{equation}

\begin{equation}
\!\!\!\boldsymbol m _{\mathrm{s}}\!\!=\!\!\!\begin{bmatrix}0.58  \!\!\!\!&0.40  \!\!\!\!\!&-0.12  \!\!\!\!&0.34  \!\!\!\!\!&-4.22  \!\!\!\!\!&-4.24  \!\!\!\!\!&-200.90\end{bmatrix}\!.
    \label{eq:16}
\end{equation}

\begin{figure*}[t]
  \centering
   \includegraphics[width=0.7\linewidth]{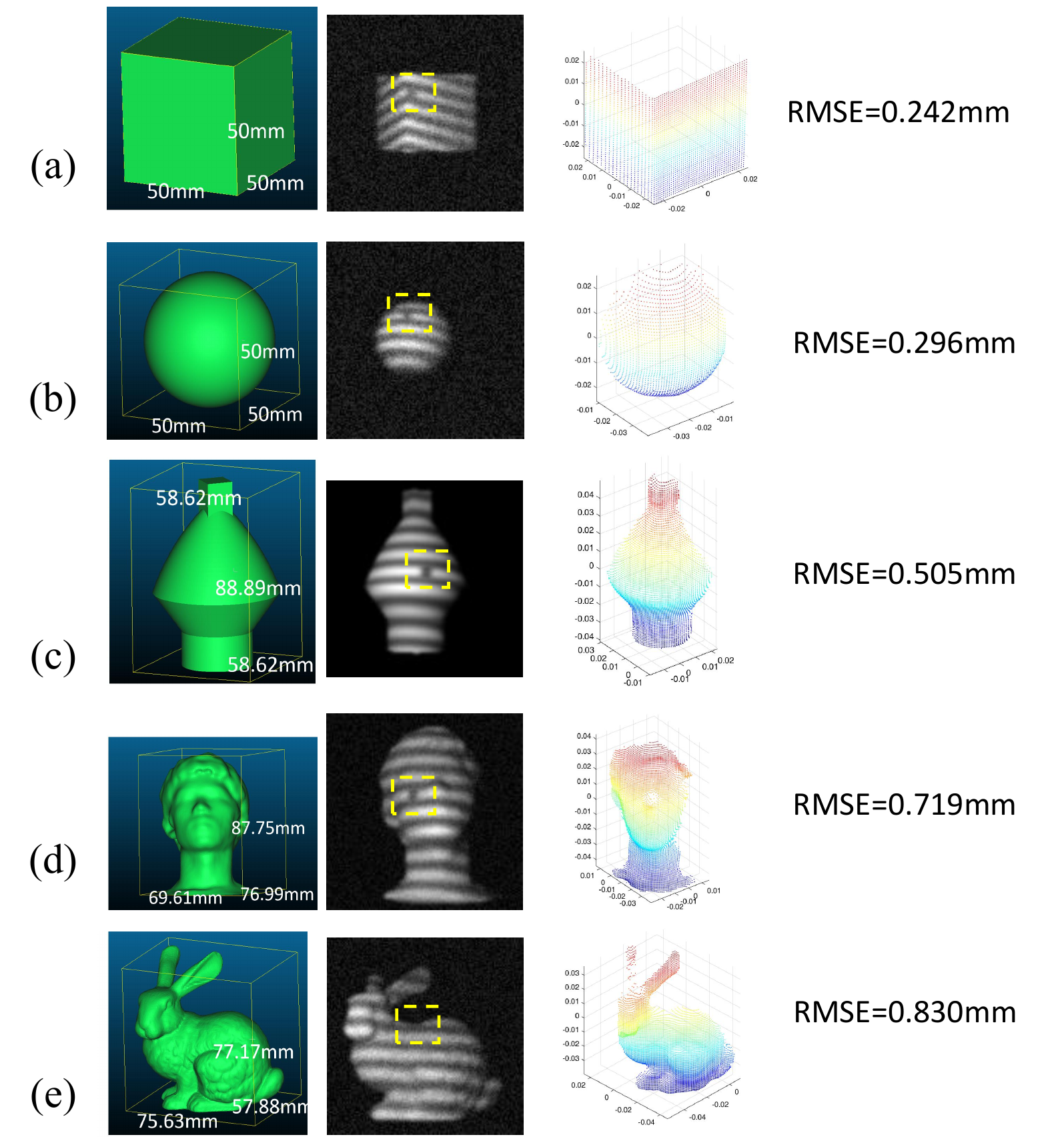}
   \caption{The simulation results of some objects, the unit of the coordinate axis is meter.}
   \label{Fig.7}
\end{figure*}

After calibration, some objects are measured, as shown in Fig. \textcolor{blue}{\ref{Fig.7}}. The yellow dashed box indicates the location of the marker. The measured point clouds are fitted with standard models, and the Root Mean Square Errors (RMSE) are recorded in Fig. \textcolor{blue}{\ref{Fig.7}}. Even for complex objects (human sculpture, bunny), the errors are less than 1mm, demonstrating the high accuracy of our method.

\subsection{Real Experiment}
\begin{figure}[t]
  \centering
   \includegraphics[width=0.8\linewidth]{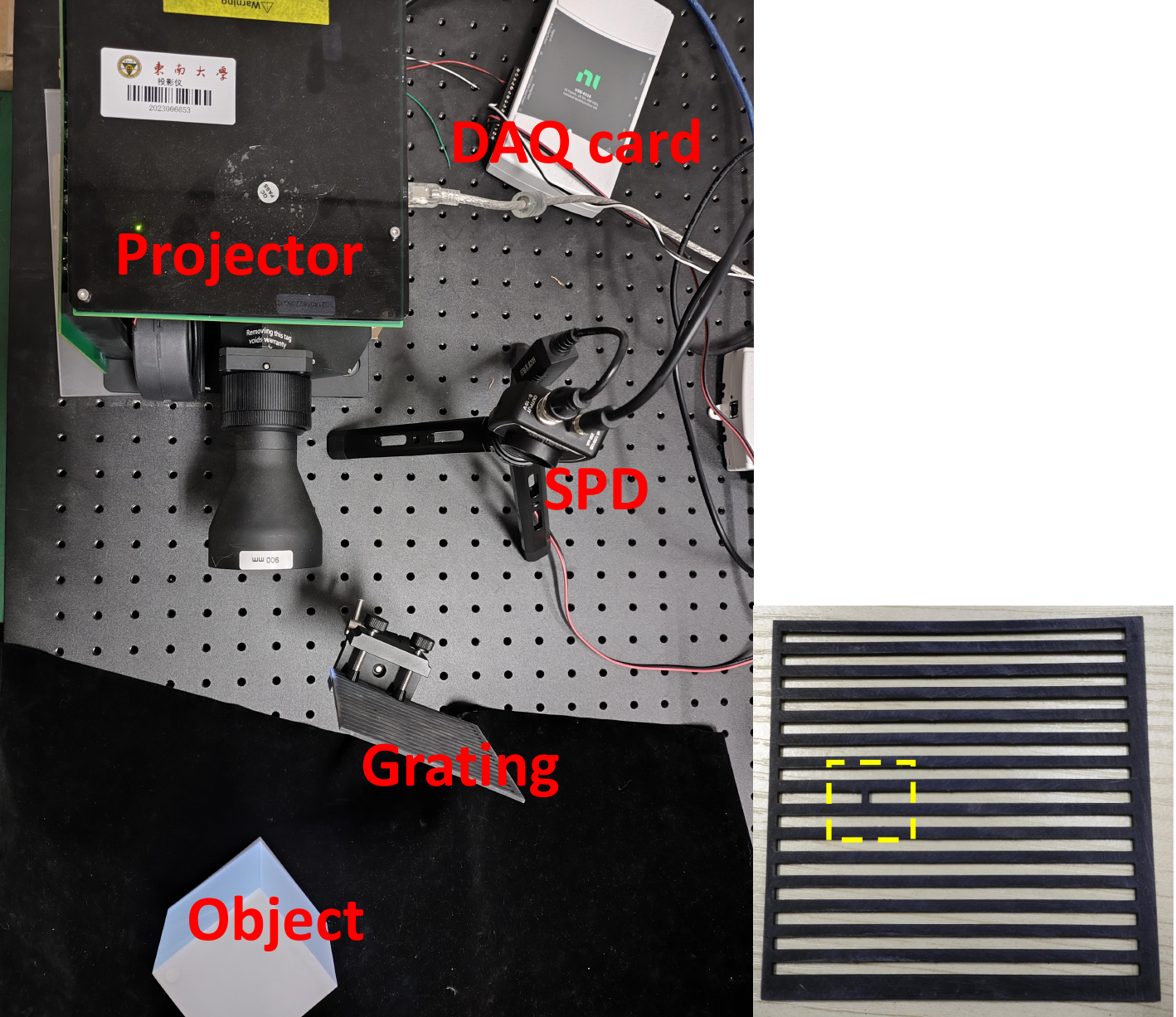}
   \caption{The experimental setup.}
   \label{Fig.8}
\end{figure}

We deploy the theoretical framework on the real 3D SPI system and verify the accuracy in practical applications. The real 3D SPI system is shown in Fig. \textcolor{blue}{\ref{Fig.8}}, where the SPD is Thorlabs PDA100A2, the data acquisition card (DAQ card) is NI USB-6216, the projector is Wintech PRO6500, and the grating is manufactured by 3D printing, with a period of 6mm and a size of 9.6cm×9.6cm. The marker used to mark the $0^\text{th}$ order is designed on the grating sheet, as indicated in the yellow dashed box in Fig. \textcolor{blue}{\ref{Fig.8}}. The method proposed in \textcolor{blue}{\cite{ref21}} is used to capture images.

\begin{figure}[t]
  \centering
   \includegraphics[width=0.8\linewidth]{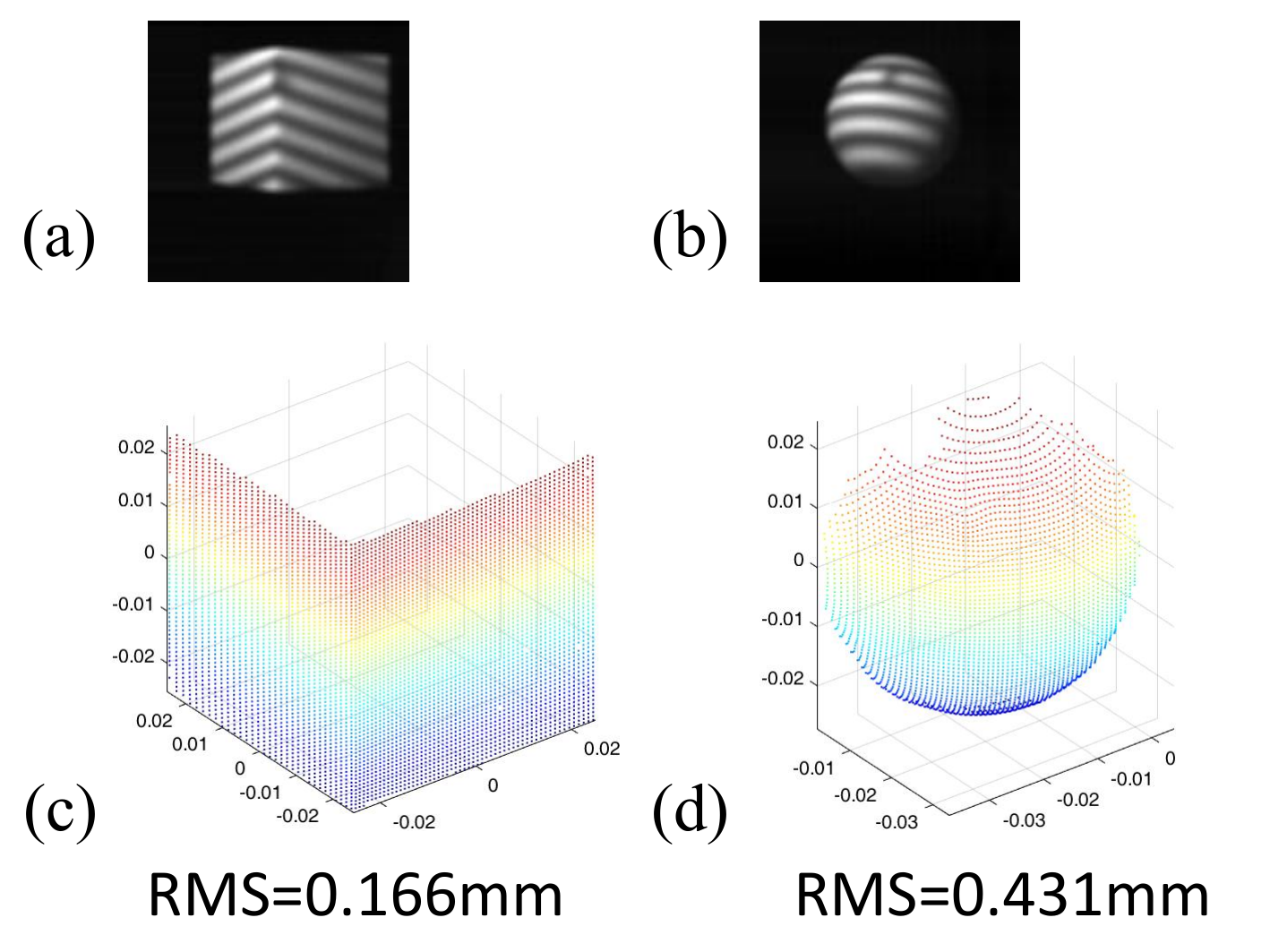}
   \caption{The results of the cube and the sphere, the unit of the coordinate axis is meter.}
   \label{Fig.9}
\end{figure}

A cube customized by 3D printing is used as the calibration object. The side length of the cube is 50.60mm. The resolution of the captured images is also 128×128. After capturing a cube fringe image, the calibration parameters are solved as
\begin{equation}
\boldsymbol M_{\mathrm{p}}=\begin{bmatrix}-0.645  &0.260  &0.001  &-0.040 \\-0.057  &-0.070  &0.674  &-0.026 \\-0.128  &-0.177  &0.003  &-0.069\end{bmatrix},
    \label{eq:17}
\end{equation}

\begin{equation}
\!\!\!\boldsymbol m_{\mathrm{s}}\!\!=\!\!\!\begin{bmatrix}0.74  \!\!\!\!&0.48  \!\!\!\!\!&-0.23  \!\!\!\!&0.25  \!\!\!\!\!&-35.43  \!\!\!\!\!&-27.03  \!\!\!\!\!&-145.04\end{bmatrix}\!.
    \label{eq:18}
\end{equation}
Then, the cube used for calibration and a sphere with a diameter of 50.14mm is measured. The reconstructed point clouds are fitted with the standard models, with the RMSE of 0.166mm and 0.431mm, respectively. The reconstructed point clouds are shown in Fig. \textcolor{blue}{\ref{Fig.9}}.

We also conduct comparisons with other state-of-the-art 3D SPI methods. Ma's work \textcolor{blue}{\cite{ref15}} and Niu's work \textcolor{blue}{\cite{ref18}} are representative methods in 3D SPI, using the phase-height method and the triangular stereo method to calibrate 3D SPI systems, respectively. Table \textcolor{blue}{\ref{tab:2}} lists the number of images needed for calibrating 3D SPI systems. We can calibrate a 3D SPI system with only one image, while a large number of images are needed in other methods. Table \textcolor{blue}{\ref{tab:3}} lists the accuracy for measuring a sphere with a diameter of about 50 mm. Our calibration object is manufactured at a lower cost and more convenience, and we use only one image for calibration or measurement, while our method achieves a similar level to other methods in terms of accuracy. Since additional devices or operations are used to assist in calibration in the other two methods, it is difficult to reproduce their methods in the virtual system or our real system. Hence, it is difficult to create the same experimental conditions to compare accuracy. Nevertheless, the above comparisons still prove that our method achieves a high level of measurement accuracy using only one image for calibration.

\begin{table}
  \centering
  \caption{The number of images needed for calibration in different methods.}
  \begin{tabular}{cccc}
    \toprule
    Method & Ma's \textcolor{blue}{\cite{ref15}} & Niu's \textcolor{blue}{\cite{ref18}} & Ours\\
    \midrule
    Number of images & 143 & 296 & \textbf{1}\\
    \bottomrule
  \end{tabular}
  \label{tab:2}
\end{table}

\begin{table}
  \centering
  \caption{The accuracy for the standard sphere in different methods.}
  \begin{tabular}{cccc}
    \toprule
    Method & \makecell[c]{Diameter\\(mm)} & Resolution & \makecell[c]{RMSE\\(mm)}\\
    \midrule
    Ma's \textcolor{blue}{\cite{ref15}} & 50.00  & 128×128 & 0.158\\
    Niu's \textcolor{blue}{\cite{ref18}} & 49.49  & 128×128 & 0.492\\
    Ours & 50.14 & 128×128 & 0.431\\
    \bottomrule
  \end{tabular}
  \label{tab:3}
\end{table}

\begin{figure}[t]
  \centering
   \includegraphics[width=1.0\linewidth]{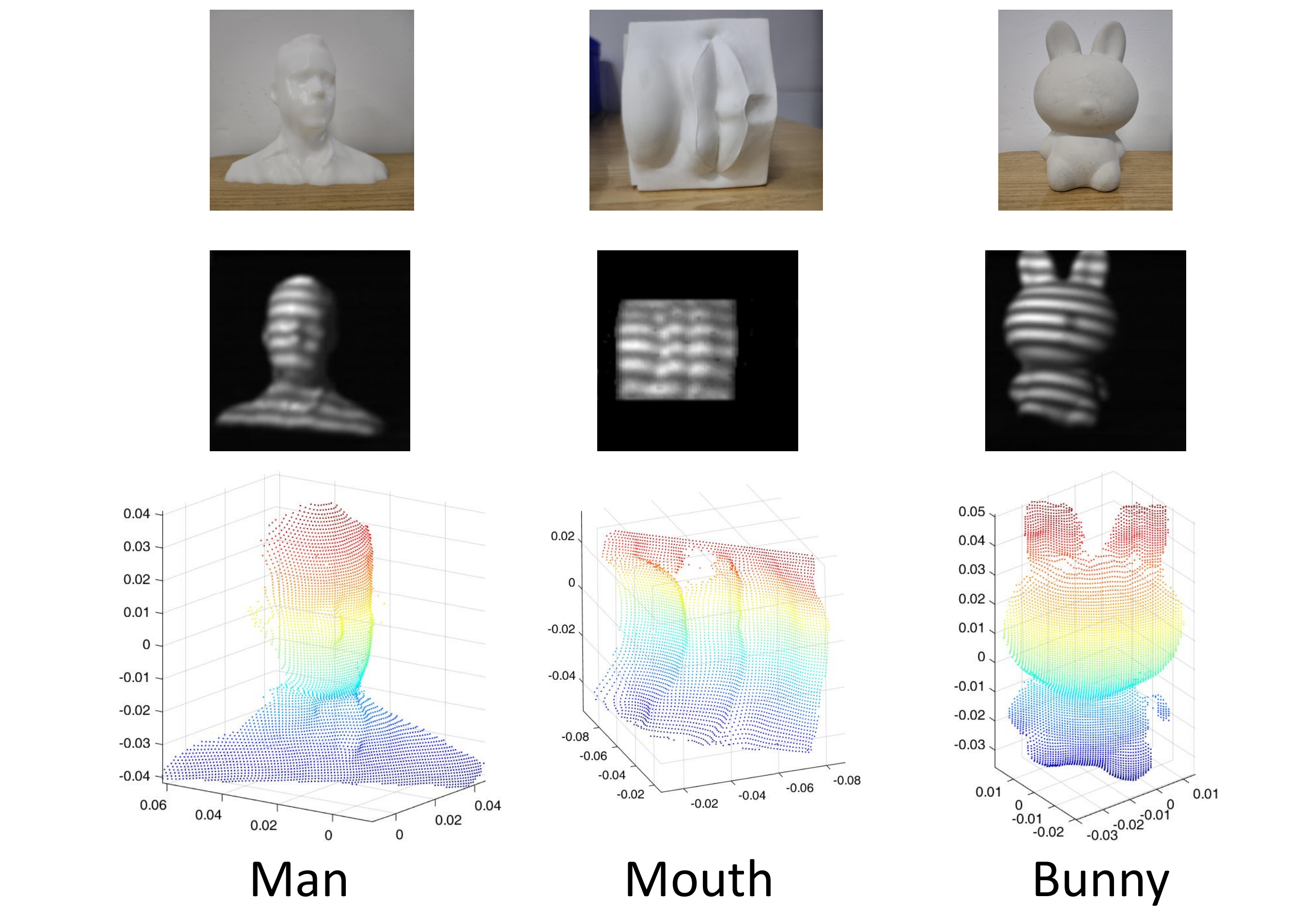}
   \caption{The results of other objects.}
   \label{Fig.10}
\end{figure}

More complex objects are measured to verify the effectiveness of our method. The results are shown in Fig. \textcolor{blue}{\ref{Fig.10}}. The first line shows the natural images of the objects, the second line shows the fringe images captured by our 3D SPI system, and the third line shows the reconstructed point clouds. It can be seen that the point clouds display the objects well, which proves the validity of our method. The point cloud of each object can be reconstructed using only one image, demonstrating the high efficiency of our method.

\section{Conclusion and Discussion}
In this paper, a general and efficient 3D SPI framework is proposed. A CaliF is proposed for calibrating the 3D SPI system, which can output a pointmap containing abundant matching point pairs from only one fringe image of the calibration object. In order to build the high-precision CaliF, a virtual FPP system is built and a digital twin of the calibration object is used to obtain the rich and accurate datasets. With our CaliF, high-precision calibration is achieved using only one image, and the experiments verify the accuracy of our new method.

Our work paves the way for more streamlined and efficient calibration processes in 3D SPI. There are many aspects to be explored further. In this research, the non-linear distortion is not considered in the calibration. For the SPDG, the distortion model cannot be constructed because only $v_\text{s}$ can be obtained using the grating sheet. For the projector, we perform an additional calibration for the real experiment, in which the parameters of the non-linear distortion $k_1$, $k_2$, $p_1$, and $p_2$ are introduced. After calibration, these four parameters are -0.00013, -0.00003, -0.00064, and 0.00018, respectively, which are small enough to demonstrate that the non-linear distortion is insignificant in our projector. Further, the accuracy of the real experiment based on the linear model is close to that of the simulation experiment which represents the upper limit of accuracy, proving that the linear model is reliable enough.

\section*{Acknowledgments}
The authors would like to thank the anonymous reviewers and editors for their profound comments.

\newpage


\begin{IEEEbiography}[{\includegraphics[width=1in,height=1.25in,clip,keepaspectratio]{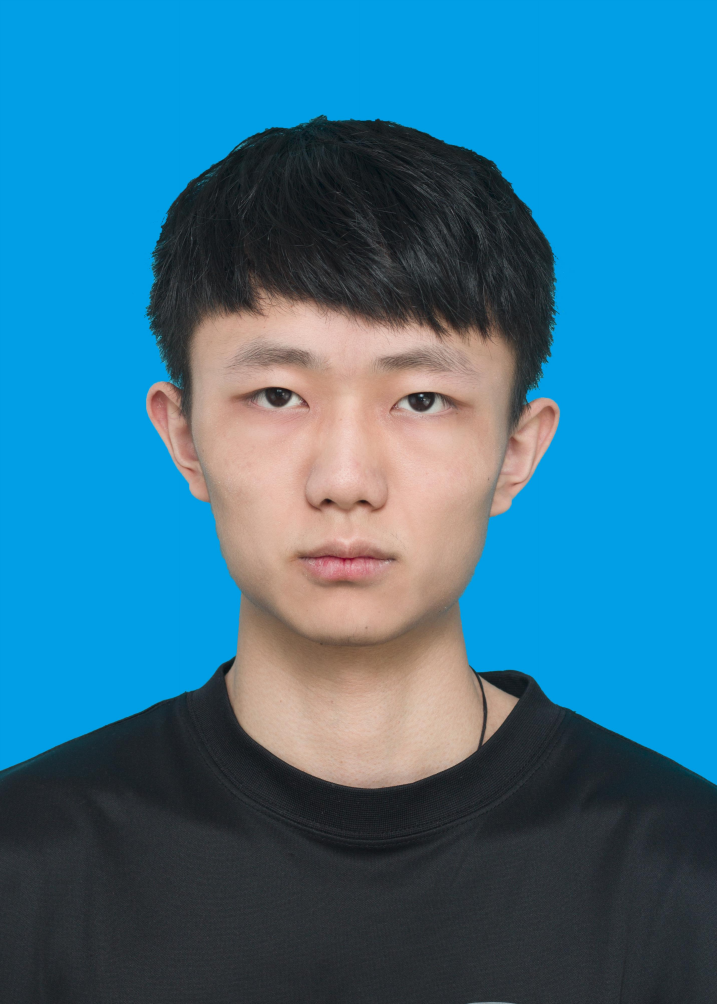}}]{Xinyue Ma}
received the bachelor’s degree in automation from Hefei University of Technology, Hefei, China, in 2020.

He is currently pursuing the Ph.D. degree with the School of Automation, Southeast University, Nanjing, China.

His research interests include single-pixel imaging and 3D measurement.
\end{IEEEbiography}

\begin{IEEEbiography}[{\includegraphics[width=1in,height=1.25in,clip,keepaspectratio]{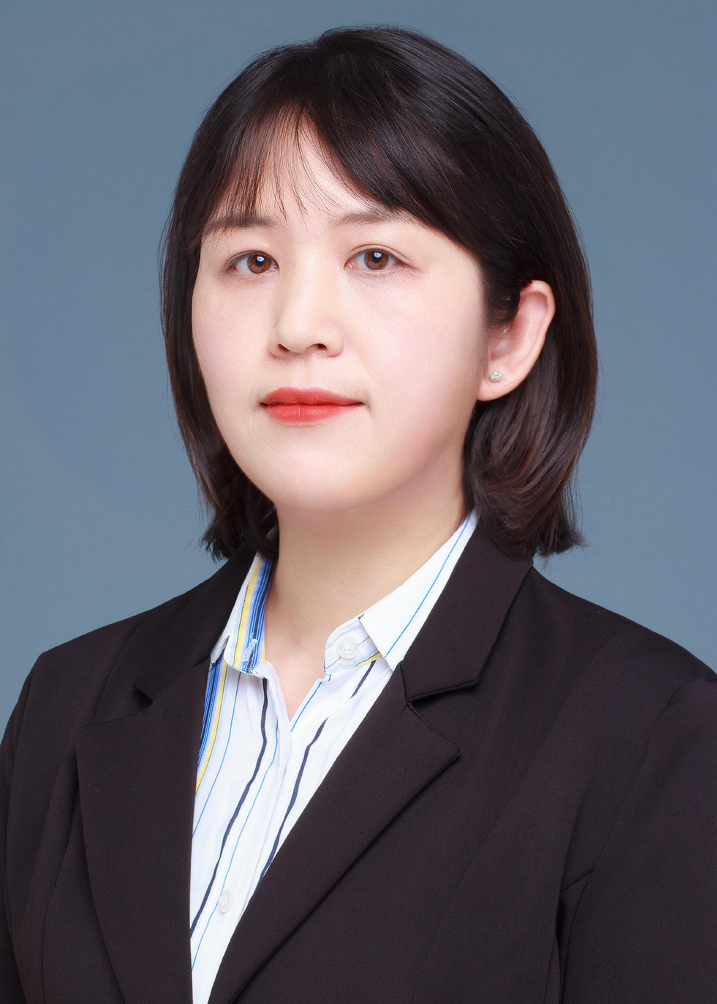}}]{Chenxing Wang}
(Member, IEEE) received her Ph.D. degree from Southeast University in 2013, and was a research fellow at Nanyang Technological University in Singapore from 2014 to 2016.

She is now an associate professor at Southeast University. Her research interests include optical 3D measurement, optical computational imaging, human 3D reconstruction, advanced signal processing, precision engineering, and automation.
\end{IEEEbiography}
\vspace{11pt}

\vfill

\end{document}